\documentclass[
    ,final            
    ,numberedheadings 
    ,cmfonts          
  ]
  {aipproc}
\usepackage{graphicx,epsfig,slashed,bm,pstricks}

\layoutstyle{6x9}

\begin{document}

\title{ Pions in the quark matter phase diagram}


\author{D.~Zablocki\thanks{Present address: Bogoliubov Laboratory for
Theoretical Physics, JINR, 141980 Dubna, Russia}\hspace{2mm}$^,$}{
   address={Instytut Fizyki Teoretycznej,
Uniwersytet Wroc{\l}awski, 50-204 Wroc{\l}aw, Poland},
altaddress={Institut f\"ur Physik,
Universit\"at Rostock, D-18051 Rostock, Germany}
}

\author{D.~Blaschke}{
  address={Instytut Fizyki Teoretycznej,
Uniwersytet Wroc{\l}awski, 50-204 Wroc{\l}aw, Poland},
altaddress={Bogoliubov Laboratory for Theoretical Physics,
JINR, 141980 Dubna, Russia}
}

\author{R.~Anglani}{
  address={Dipartimento di Fisica,
Universit\`a di Bari, I-70126 Bari, Italia},
altaddress={I.N.F.N., Sezione di Bari, I-70126 Bari, Italia \\
E-mail: zablocki@ift.uni.wroc.pl, blaschke@ift.uni.wroc.pl, roberto.anglani@ba.infn.it}
}
\begin{abstract}
The relationship between mesonic correlations and quantum
condensates in the quark matter phase diagram is explored within a quantum
field theoretical approach of the Nambu and Jona-Lasinio (NJL) type.
Mean-field values in the scalar meson and diquark channels are order
parameters signalling the occurrence of quark condensates, entailing
chiral symmetry breaking ($\chi$SB) and color superconductivity (2SC) in
quark matter.
We investigate the spectral properties of scalar and pseudoscalar meson
excitations in the phase diagram in Gaussian approximation and show that
outside the $\chi$SB region where the pion is a zero-width bound state,
there are two regions where it can be considered as a quasi-bound state with
a lifetime exceeding that of a typical heavy-ion collision fireball:
(A) the high-temperature  $\chi$SB crossover region at low densities and
(B) the high-density color superconducting phase at temperatures below 100 MeV.
\end{abstract}

\maketitle


\section{Introduction}

The study of the QCD phase diagram is a key issue in modern
theoretical and experimental physics of dense matter. Recent
heavy-ion collision experiments at RHIC Brookhaven
\cite{Muller:2006ee} have led to the insight that the quark-gluon
plasma (QGP) at high temperatures behaves as a perfect fluid with a
low viscosity to entropy ratio $\eta/s \approx 0.1 - 0.2$
\cite{Shuryak:2003xe,Shuryak:2003ty,Teaney:2004qa} which is very
close to the KSS bound \cite{KSS} for this number, $1/(4\pi)$.
This strong deviation from the behavior of a gas of weakly
interacting quarks and gluons is attributed to the occurence of
mesonic bound states
\cite{Shuryak:2003xe,Shuryak:2003ty,Shuryak:2004tx} or resonances
\cite{Blaschke:2003ut,Blaschke:2005cm,Abuki:2006dv} in the strongly
coupled QGP  (sQGP). It has been pointed out \cite{Shuryak:2006ap}
that this situation in hot and dense QCD matter bears similarities
with strongly coupled plasmas in  other systems where bound state
dissociation or Mott-Anderson delocalization \cite{Mott} occurs
since the effective coupling strength is modified by electronic
screening and/or Pauli blocking effects. It is thus a very general
effect expected to occur in a wide variety of dense
Fermi systems with attractive interactions 
\cite{Armen:2006,Bronold:2006,Redmer:2006,Schmidt:1990,Stein:1995,Schnell:1999tu,Kitazawa:2001ft,Kitazawa:2003cs,Blaschke:2004cs,Blaschke:2005uj}.
When this Mott transition from truly bound to resonantly paired states occurs
under conditions of Bose condensation one speaks of a BEC-BCS crossover
\cite{chenfig,Calzetta:2006,Gurarie:2007}.
Recently, this transition became accessible to laboratory experiments
with ultracold gases of fermionic atoms coupled via Feshbach resonances
with a strength tunable by applying external magnetic fields
\cite{Gurarie:2007,greiner:2003,Zwierlein:2003,Zwierlein:2005,Greiner:2003a}.
The BEC-BCS crossover transition in quark matter is of particular theoretical
interest due to the additional relativistic regime it offers
\cite{Abuki:2006dv,Deng:2006ed,Sun:2007fc}.

Theoretical concepts explaining the appearance of nonperturbative phenomena
like quantum condensates and bound states in dense Fermi systems with their
observable consequences shall apply here but must be formulated within a
quantum field theoretic approach.
A systematic treatment of these effects is possible within the path integral
formulation for finite-temperature quantum field theories.
This approach is especially suited
to take into account the effects of spontaneous 
symmetry breaking. Here we will apply this approach on the example
of a model field theory of the NJL type to quark matter as a
relativistic strongly interacting Fermi system. It is our aim to
delineate relationships between the regions of the NJL model phase
diagram where $\chi$SB and color superconductivity occur and the
possibility to observe quasi-bound pionic states in HIC experiments
 where  hot, dense QCD matter produced in the form of rather
short-lived fireballs.

\section{Scalar-pseudoscalar mesons in a superconducting two-flavor NJL model}

As a generic model system for the description of hot, dense Fermi-systems with
strong, short-range interactions we consider quark matter described by a
model Lagrangian with four-fermion coupling.
The key quantity is the partition function $\mathcal{Z}$ from which all
thermodynamic quantities can be derived.
In the imaginary time formalism (${t=-i\tau}$) it can be expressed
as a path integral \cite{Kapusta:1989}
\begin{equation}
\mathcal{Z} =
\int \mathcal{D}(iq^\dagger)\mathcal{D}(q)\;
e^{\int^\beta \,d^4x\,(\mathcal{L}-\mu q^{\dagger}q)}~,
\label{LL}
\end{equation}
where the chemical potential $\mu$ is introduced as a Lagrange multiplier for
assuring conservation of baryon number.
The notation $\int^\beta\, d^4x$ is shorthand for
$\int_0^\beta\,d\tau \int\,d^3x$ where $\beta$ is the inverse
temperature.
The quark matter is described by a Dirac Lagrangian with internal degrees of
freedom ($N_f = 2$ flavors , $N_c = 3$ colors), with a current-current-type
four-fermion interaction inspired by one-gluon exchange.
After Fierz transformation of the interaction, we select the scalar diquark
channel and the scalar-pseudoscalar meson channels so that our model
Lagrangian assumes the form
\begin{eqnarray}
\label{L}
\mathcal{L} = \bar{q}(  i\slashed\partial - m_0)q
+ G_D (\bar{q} i\gamma_5 C \tau_2 \lambda_2 \bar{q}^T )
       (q^T iC\gamma_5 \tau_2 \lambda_2  q)
+ G_S\big[\left(\bar{q}q\right)^2
+\left(\bar{q}i\gamma_5{\bf \tau}q\right)^2\big] ~.
\end{eqnarray}
Here $\gamma_\nu$ are the Dirac matrices,
$\lambda_2$ is a  color $SU(3)$ Gell-Mann matrix,
$\tau_i$ are $SU(2)$ flavor matrices
and $C = i\gamma^2\gamma^0$ is the charge conjugation matrix.
$G_S$ and $G_D$ are coupling strengths corresponding to
the different channels, see Ref.~\cite{Buballa:2003qv} for a recent review.
For the numerical analysis we adopt parameters from Ref.
\cite{Grigorian:2006qe} and consider $G_D$  as a free parameter.

\begin{figure}
\includegraphics[width=0.40\textwidth,angle=-90]{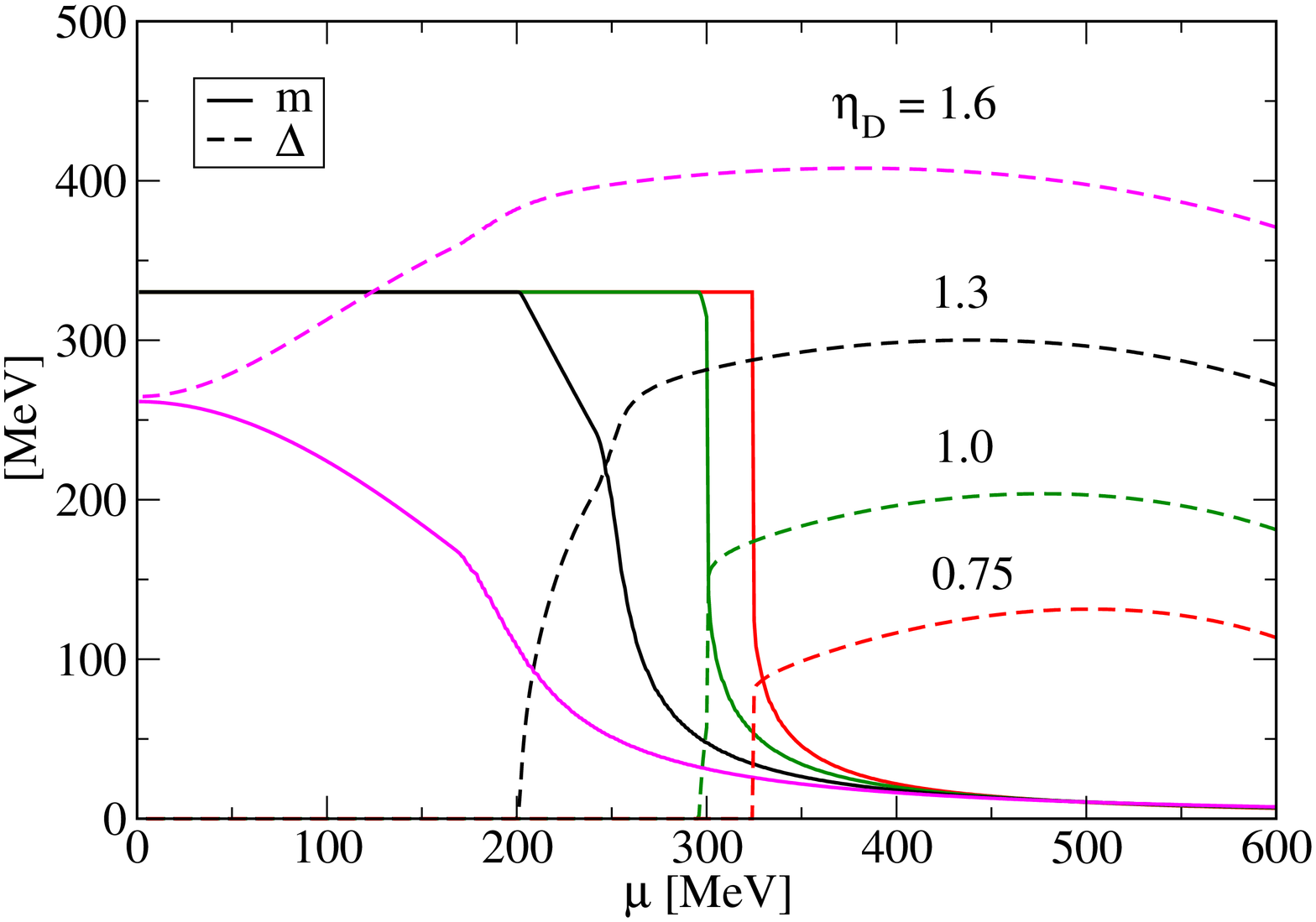}\includegraphics[width=0.40\textwidth,angle=-90]{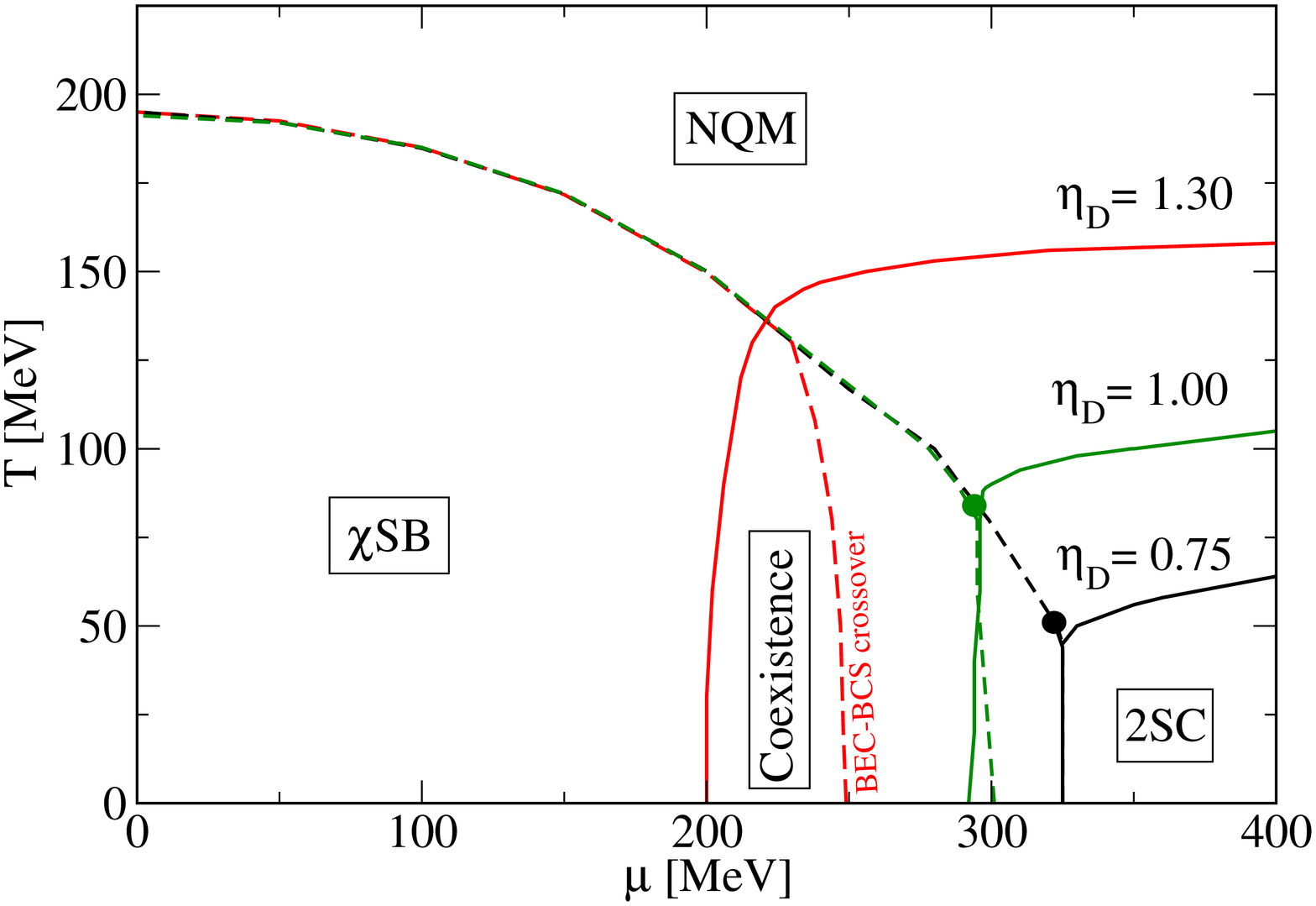}
\caption{{\it Left panel}: order parameters for $\chi$SB (full lines) and
color superconductivity (dashed lines) at $T = 0$ for
different values of $\eta_D$. First order phase
transitions turn to second order or even crossover when $\eta_D$ is increased.
For details, see text. {\it Right panel}: phase diagram of two-flavor quark matter with critical lines for chiral symmetry breaking (dashed) and color superconductivity (solid)
for three values of the diquark coupling strength: $\eta_D=0.75$, $1.0$ and $1.3$. The dots indicate the
critical endpoint for first order phase transitions.
\label{order}}
\end{figure}

After introduction of the Hubbard-Stratonovich
\cite{Kleinert:1977tv} auxiliary fields $\Delta(\tau,x)$,
$\Delta^*(\tau,x)$, $\pi(\tau, x)$,
$\sigma(\tau,x)$ and the Nambu-Gorkov spinors
$\Psi=\frac{1}{\sqrt{2}}(q q^{c})^{T}$,
$\bar{\Psi}=\frac{1}{\sqrt{2}}(\bar{q} \bar{q}^{c})$
with $q^c (x)\equiv C \bar{q}^T (x)$, the partition function becomes a Gaussian path integral in the bispinor fields which can be evaluated and yields the fermion determinant
\begin{eqnarray}
\mathcal{Z}
&=&
\int\mathcal{D}\Delta^*\mathcal{D}\Delta\mathcal{D}\sigma\mathcal{D}\pi
e^{-\int^\beta d^4x\frac{\sigma^2 + \pi^2}{4G_S}
+\frac{|\Delta|^2}{4G_D}}
\cdot
{\rm Det}[S^{-1}]~,
\label{detglei}
\end{eqnarray}
where the inverse bispinor propagator is a matrix in Nambu-Gorkov-, Dirac-,
color- and flavor space.
After Fourier transformation it reads
\begin{eqnarray}
\label{inv-prop}
S^{-1}
&=&
\left(
\begin{array}{cc}
(i\omega_n+\mu)\gamma_0 - m-i{\bm \gamma}\cdot{\bf p}
-i\gamma_5{\bm \tau}\cdot{\bm \pi}
& \hspace{-1.5cm} i\Delta\gamma_5\tau_2\lambda_2 \\
 \hspace{-1.5cm} i\Delta^*\gamma_5\tau_2\lambda_2
& \hspace{-2cm} (i\omega_n-\mu)\gamma_0
- m-i{\bm \gamma}\cdot{\bf p} +i\gamma_5{\bm \tau}^{t}\cdot{\bm \pi}
\end{array}
\right)~,
\end{eqnarray}
with $ m = m_0 + \sigma$.
So far we could derive with (\ref{detglei}) a very compact, bosonized form of
the quark matter partition function (\ref{LL}) which is an exact
transformation of (\ref{LL}), now formulated in terms of collective,
bosonic fields.
As we will demonstrate in the following, this form is suitable since it allows
to obtain nonperturbative results already in the lowest orders with respect to
an expansion around the stationary values of these fields. In performing this
expansion, we may factorize the partition function into mean field (MF),
Gaussian fluctuation (Gauss) and residual (res) contributions
$$
  Z(\mu,T)\equiv e^{-\beta\Omega(\mu,T)}
=Z_{MF}(\mu,T) Z_{\rm Gauss}(\mu,T)Z_{\rm res}(\mu,T)~.
$$
In the following we will discuss the physical content of these approximations.
In thermodynamical equilibrium, the mean field values satisfy the
stationarity condition that the thermodynamical potential
$\Omega_{MF}\equiv-\frac{1}{\beta V}\ln\mathcal{Z}_{MF}$ be minimal,
i.e. $ \frac{\partial\Omega_{MF}}{\partial\sigma_{MF}}=
\frac{\partial\Omega_{MF}}{\partial\pi_{MF}}=
\frac{\partial\Omega_{MF}}{\partial\Delta_{MF}}=0~.
\label{Ex-OmegaMF} $ This is equivalent to the fulfillment of the
gap equations $\sigma_{MF}=-4G_S{\rm Tr}\left(S_{MF}\right)\equiv
m-m_0$, ${\bm \pi}_{MF}=-4iG_S{\rm Tr}\left(\gamma_5{\bm
\tau}S_{MF}\right) =0$ and $\Delta_{MF}=4G_D{\rm
Tr}\left(\gamma_5\tau_2 \lambda_2 S_{MF}\right) =\Delta$, together
with the stability criterion that the determinant of the curvature
matrix formed by the second derivatives is positive. After the
evaluation of the traces
and the sum over
the Matsubara frequencies one gets
\begin{eqnarray}
\label{OmegaMF}
\Omega_{MF}
&=&
-\frac{1}{\beta V}\ln\mathcal{Z}_{MF}
=
\frac{(m - m_0)^2}{4G_S} + \frac{|\Delta|^2}{4G_D}
- \frac{1}{\beta V}{\rm Tr}\left(\ln\beta S_{MF}^{-1}\right)
\nonumber\\
&=&\hspace{-2mm}
\frac{(m - m_0)^2}{4G_S} + \frac{|\Delta|^2}{4G_D} -4\int\frac
{d^3p}{(2\pi)^3}\left[
E_{\bf p}^+ + E_{\bf p}^- + E_{\bf p} + 2T\ln(1+e^{-\beta E_{\bf p}^+})
\right.
\nonumber\\
&&\hspace{-2mm}
+ \left. 2T\ln(1+e^{-\beta E_{\bf p}^-})
+ T\ln(1+e^{-\beta \xi_{\bf p}^+})
+ T\ln(1+e^{-\beta \xi_{\bf p}^-})\right]~,
\label{thermodynamic-potential-MF}
\end{eqnarray}
where we have defined the particle dispersion relation
$E_{\bf p}^\pm = \sqrt{\left(\xi_{\bf p}^\pm \right)^2 + \Delta^2}$ with
$\xi_{\bf p}^\pm = E_{\bf p}\pm\mu$, $E_{\bf p} = \sqrt{m^2+{\bf p}^2}$.
The $\Delta\neq 0$ dispersion law $E_{\bf p}^{-}$($E_{\bf p}^+$) is associated
to the red and green quarks (antiquarks), whereas the ungapped
blue quarks (antiquarks) have the dispersion
$\xi^-_{\bf p}$ ($\xi^+_{\bf p}$).
With the aformentioned stationarity conditions applied to Eq.~(\ref{OmegaMF})
we obtain the gap
equations for the order parameters $m$ and $\Delta$, which have to be
solved self-consistently,
\begin{eqnarray}
m-m_0 &=& 8G_{S}m\,\int\frac{d^3p}{(2\pi)^3}\frac{1}{E_{\bf p}}
\bigg\{
\left[ 1 - 2n_F(E_{\bf p}^-) \right]\frac{\xi_{\bf p}^-}{E_{\bf p}^-}
\nonumber \\ && \hspace{1.5cm}
+\left[ 1 - 2n_F(E_{\bf p}^+) \right]\frac{\xi_{\bf p}^+}{E_{\bf p}^+}
+ n_F(-\xi_{\bf p}^+) -  n_F(\xi_{\bf p}^-)\bigg\}~,
\\
\Delta &=& 8G_{D} \int\frac{d^3p}{(2\pi)^3}\left[
\frac{1 - 2n_F(E_{\bf p}^-)}{E_{\bf p}^-}
+ \frac{1 - 2n_F(E_{\bf p}^+)}{E_{\bf p}^+}
\right]~.
\end{eqnarray}
The Fermi distribution is
$n_F(E) = (1+e^{\beta E})^{-1}$.
Solutions of the gap equations for the dynamically generated quark mass $m$
and for the diquark pairing gap $\Delta$ at $T = 0$ as a function of $\mu$
are shown in the left panel of Fig.~\ref{order}.
From the solutions 
for the order parameters
in dependence of the thermodynamical variables $T$ and $\mu$ we have
constructed the phase diagram of the present quark matter model in the
$T-\mu$ plane, see right panel of Fig.~\ref{order}.
The two order parameters, that are indicators of phase transitions,
 allow to distinguish 4 phases:
 $\Delta=0$, $m\sim m_0$: normal phase (NQM);
$\Delta\neq 0$, $m\sim m_0$: color superconductor (2SC);
$\Delta=0$, $m \gg m_0$: chiral symmetry broken phase ($\chi$SB);
$\Delta\neq 0$, $m\gg m_0$: coexistence of $\chi$SB and 2SC phases.

Increasing the diquark coupling $\eta_D=G_D/G_S$ leads to an
increase of the diquark gap and therefore a rise in the critical
temperature for the second order transition to a normal quark matter
phase. It shifts also the border between color superconductivity
(2SC) and chiral symmetry broken phase ($\chi$SB) to lower values of
the chemical potential. For very strong coupling $\eta_D\sim 1$, a
coexistence region developes where both order parameters are
simultaneously nonvanishing. Since the phase border is not of first
order, no critical endpoint can be identified in this case.
In the $\chi$SB phase pion and diquark exist as zero width bound
states. At the chiral symmetry restoration transition, they merge
the continuum of unbound states and turn into (resonant) scattering
states.
When this Mott transition occurs within the 2SC phase 
we speak of a BEC-BCS crossover.

Now we discuss the interesting question of the quasiparticle
excitations in these phases. To this end, we will expand the action
functional in the partition function
up to quadratic order in the mesonic fields and arrive at a
tractable approximation for the bosonized quark matter model.

Let us expand now the mesonic fields around their mean field values.
In these lectures, we will focus on fluctuations in the mesonic channels,
where the pion and the sigma meson will emerge as quasiparticle degrees of
freedom.
On the example of the pion we will explain the physics of the Mott transition.
The detailed investigation of the quantized diquark fluctuations, which are
also a prerequisite of the formation of baryons, will be given elsewhere
\cite{Sun:2007fc,Ebert:2004dr,Anglani:2008}.

As mentioned above, the pion mean field vanishes, but the sigma
field can be separated into mean field and fluctuations as
$\sigma\rightarrow\sigma_{MF}+\sigma$. Hence it is possible to
decompose the inverse propagator $S^{-1}$ into a mean field part and
a fluctuation part $S^{-1}=S^{-1}_{MF}+\Sigma\,$, so that in the
Gaussian approximation the fermion determinant becomes
\begin{eqnarray}
\frac{{\rm Det} \left[S^{-1}\right]\big|_{\rm Gauss}}
{{\rm Det} \left[S^{-1}_{MF}\right]}
=
\exp\left\{-\frac{1}{2}\int\frac{d^{4}q}{(2\pi)^4}
{\rm Tr}\left[S_{MF}(p)\Sigma(q) S_{MF}(p+q)\Sigma(q)\right]\right\}~,
\label{expansion-propagator}
\end{eqnarray}
where the matrix $\Sigma$ and the propagator $S_{MF}$ are defined as
\begin{eqnarray}
\Sigma\equiv\left(
\begin{array}{cc}
-\sigma - i\gamma_5{\bm \tau}\cdot{\bm \pi} & 0 \\
0  & -\sigma - i\gamma_5{\bm \tau}^{t}\cdot{\bm \pi}
\end{array}
\right)\,,\;\;\;\;\;\;S_{MF}\equiv
\left(\begin{array}{cc}
{\bf G}^+ & {\bf F}^- \\
{\bf F}^+ & {\bf G}^-
\end{array}\right)~.
\end{eqnarray}
The matrix elements of the Nambu-Gorkov propagator are
\begin{eqnarray}
{\bf G}^{\pm}_{p}
&=&
\sum_{s_p}\sum_{t_p}\frac{t_p}{2E_{\bf p}^{\pm s_p}}
\frac{t_pE_{\bf p}^{\pm s_p}-s_p\xi_{\bf p}^{\pm s_p}}
       {p_0-t_pE_{\bf p}^{\pm s_p}}
\Lambda_{\bf p}^{-s_p}\gamma_0\mathcal{P}_{\rm rg}
+\sum_{s_p}
\frac{\Lambda_{\bf p}^{-s_p}\gamma_0\mathcal{P}_{\rm b}}
       {p_{0}+s_p\xi_{\bf p}^{\pm s_p}} \,,\\
{\bf F}^\pm_p
&=&
i\sum_{s_p}\sum_{t_p}
\frac{t_p}{2 E_{\bf p}^{\pm s_p}}
\frac{\Delta^\pm}{p_0-t_pE_{\bf p}^{\pm s_p}}
\Lambda_{\bf p}^{s_p}\,\gamma_5\tau_2\lambda_2\,,
\end{eqnarray}
where $s_p,t_p=\pm1$, $(\Delta^+,\Delta^-)=(\Delta^*,\Delta)$.
For the subsequent evaluation of traces in quark-loop diagrams,
it is convenient to use this notation with projectors in color space,
$\mathcal{P}_{\rm rg}={\rm diag}(1,1,0)$,
$\mathcal{P}_{\rm b}={\rm diag}(0,0,1)$ and in Dirac space,
$\Lambda^\pm_{\bf p}=
\frac{1}{2}\left[1\pm\gamma_0
\left(\frac{{\bm \gamma}\cdot{\bf p}+\hat m}{E_{\bf p}}\right)\right]\,.$
The summation over  Matsubara frequencies $p_0=i\omega_n$ is most
systematic using the above decomposition into simple poles in the $p_0$ plane.
The poles of the normal propagators $\bf G^\pm$  are given by the gapped
dispersion relations for the paired  red-green quarks (antiquarks),
$E^\pm_{\bf p} = \sqrt{(\xi^\pm_{\bf p})^2+\Delta^2}$,
and the ungapped dispersions $\xi^{\pm}_{\bf p} = E_{\bf p}\pm \mu$ for the
blue quarks (antiquarks).
The anomalous propagators $\bf F^\pm_{\bf p}$ are only nonvanishing in the
2SC phase where the pair amplitude is nonvanishing.
Let us notice explicitly that this procedure has yielded an effective action
that includes the fluctuation terms responsible for the excitation of
scalar and pseudoscalar mesonic modes.
The evaluation of the traces (\ref{expansion-propagator}) can be performed
with the result
\begin{eqnarray}
\frac{1}{2}{\rm Tr}\left(S_{MF}\Sigma S_{MF}\Sigma\right)
=
\left(
	{\bm \pi}, \sigma
\right)
\left(
	\begin{array}{cc}
		\Pi_{\pi\pi}&0\\
		0&\Pi_{\sigma\sigma}\\
	\end{array}
\right)
\left(
	\begin{array}{c}
		{\bm \pi}\\ \sigma
	\end{array}
\right)~,
\end{eqnarray}
where we have introduced the polarization functions
\begin{eqnarray}
\Pi_{\sigma\sigma}(q_0,{\bf q})
&\equiv&
{\rm Tr}[
        {\bf G}^+_p{\bf G}^+_{p+q}
        +{\bf F}^-_p{\bf F}^+_{p+q}
        +{\bf G}^-_p{\bf G}^-_{p+q}
        +{\bf F}^+_p{\bf F}^-_{p+q}
]~,
\\
\Pi_{\pi\pi}(q_0,{\bf q})
&\equiv&
-{\rm Tr}[{\bf G}^+_p(\gamma_5{\bm\tau}){\bf G}^+_{p+q}
(\gamma_5{\bm \tau})
+{\bf F}^-_{\bf p}(\gamma_5{\bm\tau}^t)
{\bf F}^+_{p+q}(\gamma_5{\bm \tau}) \nonumber \\
&&
+{\bf F}^+_p(\gamma_5{\bm \tau})
{\bf F}^-_{p+q}(\gamma_5{\bm \tau}^t)
+{\bf G}^-_p(\gamma_5{\bm \tau}^t)
{\bf G}^-_{p+q}(\gamma_5{\bm \tau}^t)]~,
\end{eqnarray}
as the key quantities for the
investigation of mesonic bound and scattering states in quark matter.
Indeed by using Bethe Salpeter equation and by evaluating the spectral
functions one can obtain important information on the mesons properties.
In the following we perform the further evaluation and discussion for the
pionic modes, the $\sigma$ mode is treated in an analogous way.
We start with the evaluation of traces and Matsubara summation and obtain
\begin{eqnarray}
&&
\Pi_{\pi\pi}(q_0,{\bf q})
=
2\int\frac{d^3p}{(2\pi)^3}\sum_{s_p,s_k}
\left(1+ s_ps_k\frac{{\bf p}\cdot({\bf p+q})-m^2}{E_{\bf p}E_{\bf p+q}}\right) \bigg\{
        \frac{n_F(s_p\xi_{\bf p}^{s_p}) - n_F(s_k\xi_{\bf p+q}^{s_k})}
               {q_0 - s_k\xi_{\bf p+q}^{s_k} + s_p\xi_{\bf p}^{s_p}}
\nonumber \\
&&\hspace{0.5cm}
+ \frac{n_F(s_p\xi_{\bf p}^{s_p})-n_F(s_k\xi_{\bf p+q}^{s_k})}
               {q_0 + s_k\xi_{\bf p+q}^{s_k} - s_p\xi_{\bf p}^{s_p}}
   +
        \sum_{t_p,t_k}
        \frac{t_pt_k}{E_{\bf p}^{s_p}E_{\bf p+q}^{s_k}}
        \frac{n_F(t_pE_{\bf p}^{s_p})-n_F(t_kE_{\bf p+q}^{s_k})}
               {q_0 - t_kE_{\bf p+q}^{s_k} + t_pE_{\bf p}^{s_p}}
\nonumber \\
&& \hspace{0.5cm}
\times \left(t_pt_kE_{\bf p}^{s_p}E_{\bf p+q}^{s_k}
+ s_ps_k\xi_{\bf p}^{s_p}\xi_{\bf p+q}^{s_k} - |\Delta|^2 \right)
\bigg\} ~.
\end{eqnarray}
For a pionic mode at rest in the medium (${\bf q}=0$) this reduces to
\begin{eqnarray}
&& \Pi_{\pi\pi}(q_0,{\bf 0}) =
8\int\frac{d^3p}{(2\pi)^3}
\bigg\{
    N(\xi_{\bf p}^+,\xi_{\bf p}^-)
    \left[\frac{1}{q_0 - 2E_{\bf p}} - \frac{1}{q_0 + 2E_{\bf p}}\right]
\nonumber \\
&& \hspace{-5mm} +
\left[1- \frac{\xi_{\bf p}^+\xi_{\bf p}^- + \Delta^2}
{E_{\bf p}^+E_{\bf p}^-}\right]
    M(E_{\bf p}^+, E_{\bf p}^-)
   \left[\frac{1}{q_0 - E_{\bf p}^+ + E_{\bf p}^-}
 - \frac{1}{q_0 + E_{\bf p}^+ - E_{\bf p}^-}\right]
\nonumber \\
&& \hspace{-5mm} +
 \left[1+\frac{\xi_{\bf p}^+\xi_{\bf p}^- +\Delta^2}{E_{\bf p}^+E_{\bf p}^-}
\right]
 N(E_{\bf p}^+, E_{\bf p}^-)
 \left[\frac{1}{q_0 - E_{\bf p}^+ - E_{\bf p}^-}
- \frac{1}{q_0 + E_{\bf p}^+ + E_{\bf p}^-}\right]
\bigg\}~,
\label{mu1}
\end{eqnarray}
where we have introduced the phase space occupation factors
$N(x,y) = 1-n_F(x)-n_F(y)$ (Pauli blocking) and $M(x,y) = n_F(x)-n_F(y)$.

We make use of the Dirac identity
$\lim_{\eta\to 0}\frac{1}{x + i\eta} = \mathcal{P}\frac{1}{x} - i\pi\delta(x)$
in order to decompose the polarization function into real and imaginary
parts after analytical continuation to the complex plane.
The imaginary part is straightforwardly integrated after transformation from
momentum to energy $\omega$.
At the pole, the integration variables transform as
$
p_\omega
=
\sqrt{\frac{\omega^4 - 4\omega^2(\mu^2 + \Delta^2)}
               {4(\omega^2 - 4\mu^2)} - m^2 }~,
$
and the integration borders shift $p\in (0,\infty) \to \omega\in
(X_\pm,\infty)$, where the thresholds are given by $2m$ for
$\Delta=0$ and $ X_\pm =
        \sqrt{(m+\mu)^2+\Delta^2}\pm\sqrt{(m-\mu)^2+\Delta^2}
\label{mu2} $ otherwise.
In this way one can decompose the pion polarization function in the 2SC phase
into a real and an imaginary part.
From the real part we can calculate the pion mass by solving the
Bethe-Salpeter equation $1-2G_\sigma{\rm Re}\Pi_{\pi\pi}(m_\pi, {\bf
0})=0$, while a nonvanishing imaginary part corresponds to a finite
pion width $\Gamma_\pi
$ for decay into
quark-antiquark pairs.

\begin{figure}[t]
\includegraphics[width=0.4\textwidth,angle=-90]{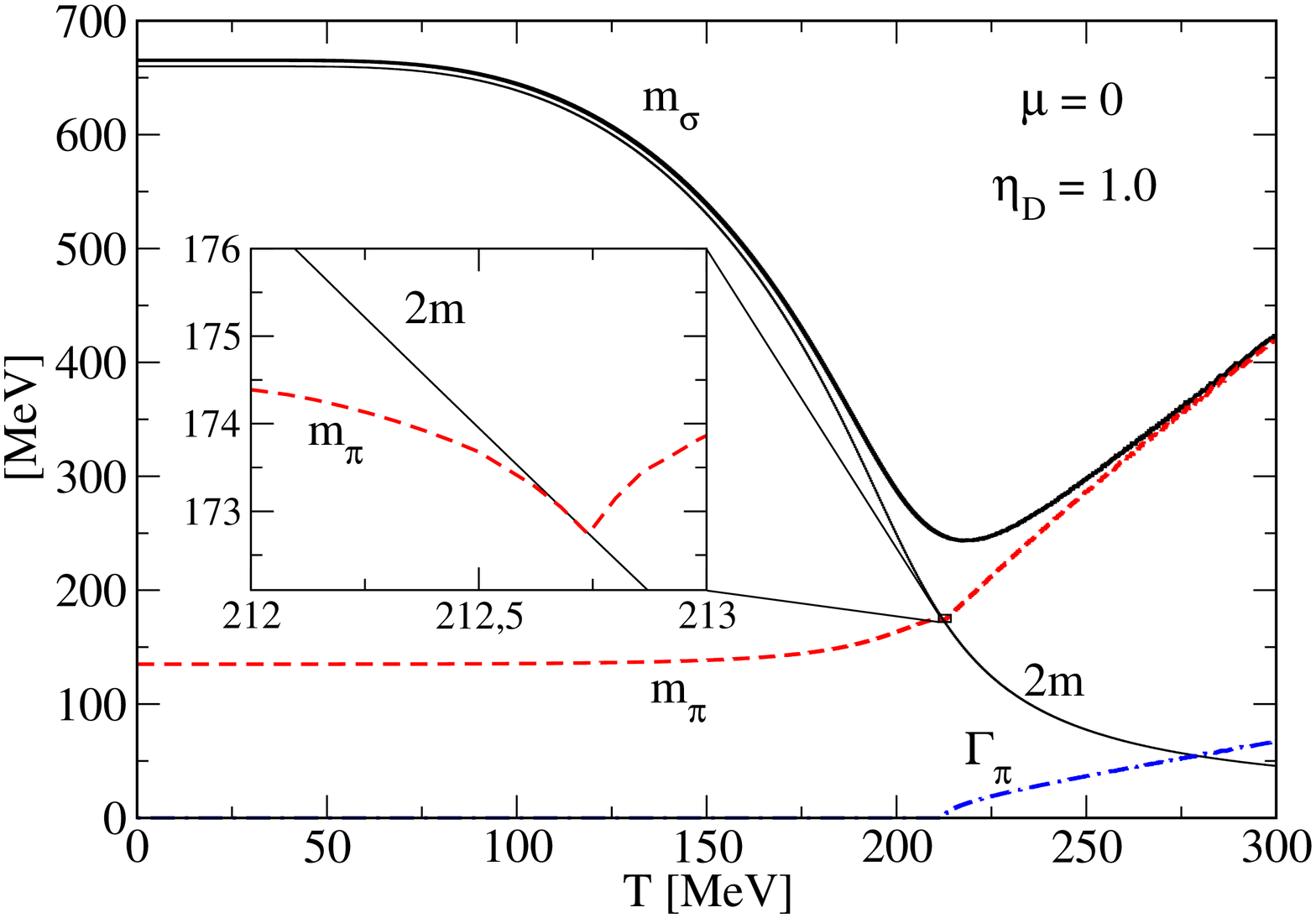}
\includegraphics[width=0.4\textwidth,angle=-90]{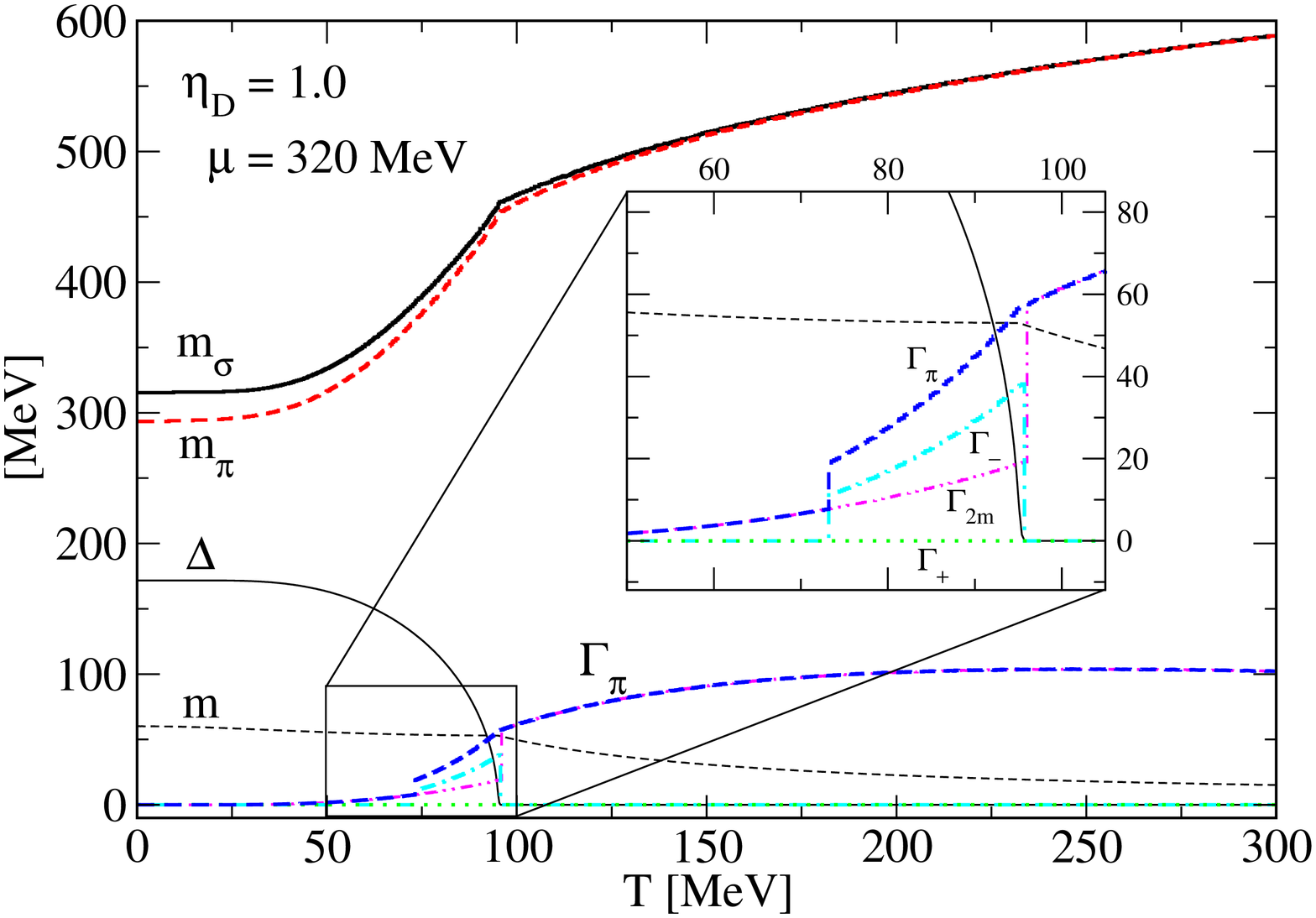}
\caption{{\it Left panel:}
Mass spectrum of mesons ($\pi$, $\sigma$) as a function
of the temperature for vanishing chemical potential $\mu_B = 0$ and strong
diquark coupling $\eta_D=1.0$.
The threshold  $E_{\rm th} = 2m_q$ for Mott dissociation of pions and
occurrence of a nonvanishing decay width $\Gamma_\pi={\rm Im}~\Pi_\pi/m_\pi$
is reached at $T_{\rm Mott} = 212.7$ MeV (see inset).
{\it Right panel:}
Mass spectrum of mesons ($\pi$, $\sigma$) as a function
of the temperature for finite chemical potential $\mu_B=320$ MeV and strong
diquark coupling $\eta_D=1.0$ in the 2SC phase.
Below the threshold  $E_{\rm th}=2~m_q$ for the onset of the decay width
$\Gamma_{2m}$ there is another process due to the lower threshold
$E_+ - E_-$ switching on (see inset).
}
\label{pion_mu000}
\end{figure}

In Fig.~\ref{pion_mu000}, we show the temperature dependence of the masses and
widths of pions and sigma-mesons for strong diquark coupling $\eta_D
= 1.0$ at vanishing chemical potential (left panel) and at  $\mu_B =
320$ MeV (right panel). At $\mu_B = 0$, the only threshold for the
imaginary parts of meson decays is  $2m$, since   $\Delta = 0$. The
$\sigma$ mass is always above the threshold and therefore this state
is unstable in the present model. The pion, however, is a bound
state until the critical temperature for the Mott transition $T_{\rm
Mott} = 212.7$ MeV is reached.
As can be seen from the slow rise of the decay width
$\Gamma_\pi$, the pion is still a well-identifyable, long-lived resonance.
The detailed analytic behavior of the pion at the Mott transition has been
discussed in the context of the NJL model by H\"ufner et al.
\cite{Hufner:1996pq}, see also the inset of the left panel of
Fig.~\ref{pion_mu000}. It shows strong similarities with the behavior of
bound states of fermionic atoms in traps when their coupling is tuned by
exploiting Feshbach resonances in an external magnetic field, see
\cite{AnnPhys}.
In the context of RHIC experiments, one has discussed such quasi-bound
states as an explanation for the perfect liquid behavior of the sQGP
\cite{Shuryak:2004tx}.

Next we want to discuss the pionic excitations in the presence of a
diquark condensate in the 2SC phase, see the right panel of
Fig.~\ref{pion_mu000}.
We observe the remarkable fact that the 2SC condensate stabilizes the pion
at $T = 0$ as a true bound state, although the pion mass exceeds by far the
threshold $2m$. This effect is due to a compensation of gapped and ungapped
quark modes and has been discussed before by Ebert et al.
 \cite{Ebert:2004dr} for $T = 0$ only. Here we extend this study to the
finite temperature case, where the pion obtains a finite width but is still
a very good resonance.
\begin{figure}[t]
\epsfig{figure=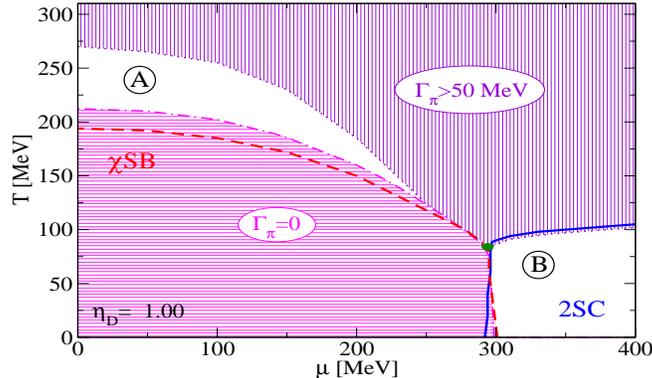,height=0.65\textwidth,width=0.43\textwidth,angle=-90}
\caption{Phase diagram of two-flavor quark matter  for the diquark coupling
strength $\eta_D=1.0$.
The pion is a stable bound state in the horizontally hatched region below the
dash-dotted line which is correlated with critical line for chiral symmetry
breaking (dashed). In the vertically hatched region above the dotted line
the pion is a short-lived resonance with a width $\Gamma_\pi > 50$ MeV
($\tau_\pi < 1$ fm/c).
There are two regions where the pion is a quasi-bound state in the quark
plasma: (A) in the crossover region above the chiral phase transition and
(B) in the color superconducting phase delimited by the solid line.
}
\label{phase_life}
\end{figure}

Finally, an interesting insight can come looking at Fig.~\ref{phase_life}.
In this figure we have reported the essential phase diagram of the two-flavor
quark matter in the case of coupling $\eta_{D}=1.00$.
The dashed line represents the chiral crossover line and the solid line refers
to the phase transition towards 2SC phase.
The dot-dashed line shows the border above which the pion
turns out to be a quasi bound state in the quark plasma. The dotted
line indicates the region border (A) below which the quasi-stable
pionic states have a lifetime greater than 1 fm/c, which is a
typical lifetime of fireballs
\footnote{Note that the physical width $\Gamma_\pi=\left(\frac{\partial {\rm Re}\Pi_{\pi\pi}}{\partial m_\pi^2}\right)^{-1}\frac{{\rm Im}\Pi_{\pi\pi}}{m_\pi}=\tau^{-1}$}. Thus they would be measured as bound
states and could be significant in the framework of HIC experiments.
Region (B) confirms our claim about the absence of stable pions in a
two-flavor superconductor. However this claim needs a comment. First
of all, it can be easily understood how the presence of a finite
diquark gap (2SC phase) at T=0 stabilizes the pion. Indeed as soon
as the quark mass drops due to chiral symmetry restoration, the
diquark gap tends to be finite and thus takes over the role of the
quark mass in the dispersion relations. Roughly speaking the quark
start to be dressed by his interactions. Thus the pion does not
``feel'' the drop of the quark mass. But as $\Delta$ melts with
increasing $T$,  the pion width $\Gamma_{\pi}$ raises, first slowly
and then rapidly, reflecting the behaviour of $\Delta$ as a function
of T. This leads inevitably to a destabilization of the the pion
states in the vicinity of the 2SC phase border. The discussion of
the mesonic modes in the 2SC phase points to a very rich spectrum of
excitations which eventually leads to specific new observable
signals of this hypothetical phase. The CBM experiment planned at
FAIR Darmstadt  and the NICA project at JINR Dubna could be capable
of creating thermodynamical conditions for the observation of these
excitations in the experiment. One promising signal could be the
scalar resonance in the pion-pion scattering at the two-pion
threshold which is in principle observable, e.g., in the two-photon
decay channel \cite{Volkov:1997dx,Blaschke:2006ss}. However, the
description of this state goes beyond the Gaussian approximation to
which we restrict ourselves in 
this work.
\section{Conclusions}

In this work we have derived and evaluated the gap equations and the
scalar-pseudoscalar meson spectra within a path integral approach to
the two-flavor NJL type model of superconducting quark matter.

After fixing the parameters of the model to the light meson spectrum
in the vacuum, the diquark coupling remains as a free parameter which
has been used to extend the model beyond the traditional range of applications
into the region of BEC-BCS crossover.

We have presented the phase diagram of quark matter at
strong and very strong coupling.
The origin of the  BEC-BCS crossover in superconducting quark matter is the
Mott transition for diquark bound states. We explain the physics of the Mott transition on the example of mesonic
correlations. We have investigated the meson spectra
(bound and scattering states) outside ($T > T_c$) and for the first time also
inside  ($T < T_c$) the color superconductivity region.
We find the thresholds for the dissociation of pionic bound states
into unbound, but resonant scattering states in the quark-\-antiquark
continuum and we have shown that
outside the $\chi$SB region where the pion is a zero-width bound state,
there are two regions where it can be considered as a quasi-bound state with
a lifetime exceeding that of a typical heavy-ion collision fireball:
(A) the high-temperature  $\chi$SB crossover region at low densities and
(B) the high-density color superconducting phase at temperatures below 100 MeV.
\begin{theacknowledgments}
D.Z. and D.B. are grateful to the organizers for providing  support
for their participation at the meeting. The work of D.B. is
supported in part by the Polish Ministry of National Education
(MENiSW). We acknowledge support from the ESF Research Networking
Programme ``CompStar'' for our participation at the workshop ``The
Complex Physics of Compact Stars''  in L\c{a}dek Zdr\'oj (Poland),
where parts of this work have been completed.
\end{theacknowledgments}

\end{document}